# LEAST SQUARE CHANNEL ESTIMATION FOR IMAGE TRANSMISSION WITH OFDM OVER FADING CHANNEL

Usha S.M[1] and Mahesh H.B[2]

[1]Department of Electronics and Communication Engineering, JSS Academy of Technical Education, Bengaluru, Karnataka, India
[2]Department of Computer science and Engineering, PES University, Bengaluru, Karnataka, India

## ABSTRACT

*Wireless communication is the most effective communication to convey audio or video information among the population. It enables the masses to connect throughout the world. Wireless technologies improve the lifestyle of individuals in rural and poor communication areas. In this view, the quality of a reliable signal can be enhanced by minimizing carrier interference. In this paper bit error rate of an image, signal is transmitted over fading channel is analyzed using orthogonal frequency multiplexing and channel estimation technique. An Orthogonal Frequency Multiplexing (OFDM) provides prominent bandwidth effectiveness and improved immunity to the fading environments. In OFDM, the data is modulated using multiple numbers of subcarriers that are orthogonal to each other. A cyclic prefix is infixed between OFDM symbols to annihilate the inter symbol interference (ISI) and inter-carrier interference (ICI). The least square channel estimation method is used to minimize the effect of multipath fading. An image, signal is modulated using BPSK, QPSK, 16QAM and 64QAM digital modulation schemes with OFDM and channel estimation and transmitted over AWGN and fading channel. The objective of this work is to improve the signal to noise ratio by reducing interference. The bit error rate vs. signal to noise ratio for BPSK, QPSK, 16QAM and 64QAM without channel estimation at 5dB is 0.4948, 0.4987, 0.4965 and 0.4983 and with channel estimation is 0.099, 0.2600, 0.3900 and 0.4300 respectively. The bit error rate obtained with BPSK, QPSK, 16QAM and 64QAM without channel estimation at SNR of 10dB is 0.4964, 0.4985, 0.4957 and 0.4982 and with channel estimation is 0.033, 0.19, 0.34 and 0.38 respectively. The bit error rate obtained with BPSK, QPSK, 16QAM and 64QAM without channel estimation at SNR of 15dB is 0.4938, 0.4985, 0.4953 and 0.4979 and with channel estimation is 0.0072, 0.1900, 0.3241 and 0.3762 respectively. The error rate is minimized with channel estimation. The error rate increases with the order of modulation and it is noticed that the error rate is minimum with the BPSK modulator and is maximum with 64QAM.*

## 1. INTRODUCTION

The communication system comprises of a transmitter, communication channel and receiver. The function of the transmitter is to modify the message signal into a structure appropriate for transmission through the channel [1-2]. This modification is achieved through modulation. The communication channel may be a transmission link, optical fiber or free space. The transmitted signal is malformed by nonlinearities or imperfections while propagation. Noise and distortion comprise two basic muddles in the conception of transmission systems. The transmitter and receiver are diligently planned to denigrate the effects of channel noise and distortion on the quality of reception [3-4]. This reconstruction is attained by an operation called demodulation. The data can be transmitted through analog or digital modulation techniques. Amplitude and angle modulation are analog modulation types. The digital modulation technique based on keying is classified as ASK, FSK (AFSK, MFSK, PSK, (BPSK, QPSK, QAM, DQPSK, OQPSK) and QAM (8QAM, 16QAM, 64QAM, 256QAM, 512QAM) [5-10].

The Double sideband suppressed carrier (DSB-SC) and single sideband suppressed carrier (SSB-SC) are used to save power and bandwidth [11-12]. The SSB-SC has the practical difficulty of isolating one sideband while transmitting the information signal. Vestigial sideband (VSB) modulation is used to overcome this difficulty. VSB-SC is a compromise between SSB-SC and DSB-SC modulation. QAM utilizes both the amplitude and frequency variations to enhance the efficiency of the radio communication system. In this work, the significance is given towards enhancing the signal to noise ratio and efficient usage of the spectral width using OFDM. First section briefs about the introduction part. In the second section of this paper, BPSK, QPSK, QAM modulator and demodulator are discussed with the block diagram [13-19]. In the third section OFDM system and methodology incorporated for image, transmissions are talked over. In the fourth section implementation and results are discussed and analyzed.

## 2. DIGITAL MODULATION TECHNIQUES

In this section, the digital modulation schemes BPSK, QPSK, 16QAM, and 64QAM modulator and demodulator are discussed with block diagrams.

### 2.1. Binary Phase shift Keying (BPSK) Modulator

The simplest form of shift keying is the BPSK. The two phases of BPSK are separated by 180°. The constellation points are positioned at 0° and 180° [20]. Therefore it handles maximum noise level and most robust of all PSK's. It is inappropriate for high data rate applications because of only able to modulate at 1 bit/symbol. Figure.1 shows the BPSK modulator. The information signal is NRZ encoded and multiplied with a high frequency carrier signal to generate BPSK modulated output. Two binary states are logic 0 and logic 1, the value of the phase angle for binary 0 is 180 and for binary 1 is 0 [21].

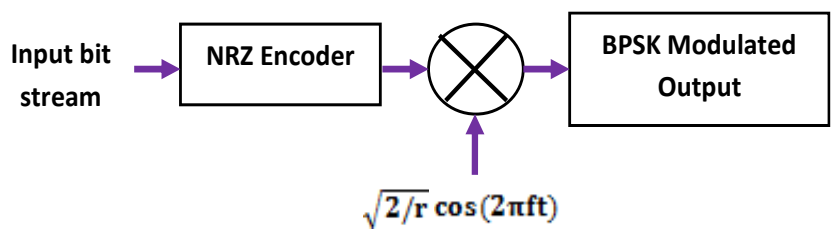

Figure 1. BPSK modulator

### 2.2. Binary Phase Shift Keying Demodulator

Figure 2 shows the BPSK demodulator block diagram. The received BPSK signal is multiplied by reference signal from the carrier recovery blocks. The multiplied output is integrated over one bit period [22]. A threshold detector attains a decision on each integrated bit based on a threshold value to determine the information bit received.

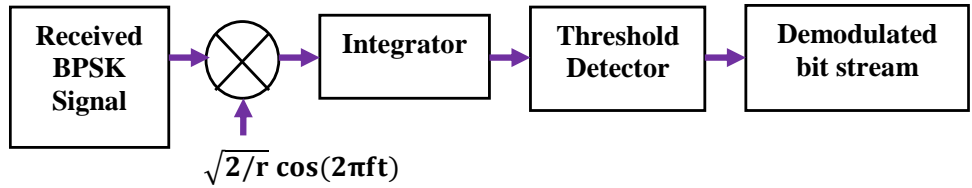

Figure 2. Binary Phase shift Keying Demodulator

## 2.3. Quadrature phase shift Keying Modulator

The QPSK expends four points on the constellation diagram. The four phases are equi-spaced across a circle. It can encode four bits/symbol and modulates two bit at once, selecting one of the four possible carrier phase shifts (0º, 90º, 180º, and 270º) [23]. The QPSK doubles the transmission capacity than BPSK using same bandwidth. QPSK employed for satellite transmission of MPEG2, video, cable modems, video conferencing, cellular phone system and further contours of digital communication over RF carrier. Figure 3 depicts the conceptual transmitter configuration for QPSK. The binary information sequence is split into in-phase and quadrature-phase portions. These are then individually modulated onto two orthogonal basis functions. In-phase and quadrature signals are imparted to generate the QPSK signal.

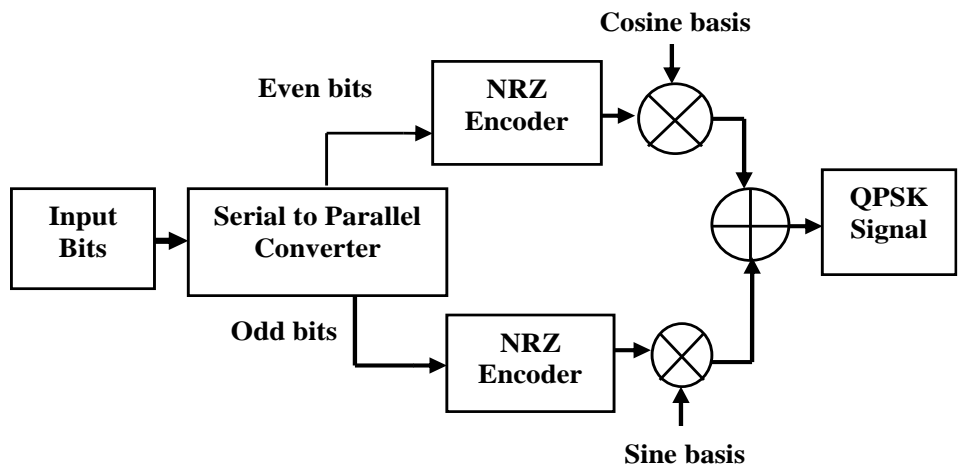

Figure 3. Quadrature Phase shift Keying Modulator

## 2.4. Quadrature Phase shift Keying Demodulator

The QPSK demodulator employs two product demodulator circuits with local oscillator, low pass filter, threshold detector, bit jointer and decoder as shown in Figure 4. The product detector simultaneously demodulates the two BPSK signals. The duad of bits is recuperated from the original data. The threshold detector utilizes a reference threshold assessment to determine logic 1 or logic 0.

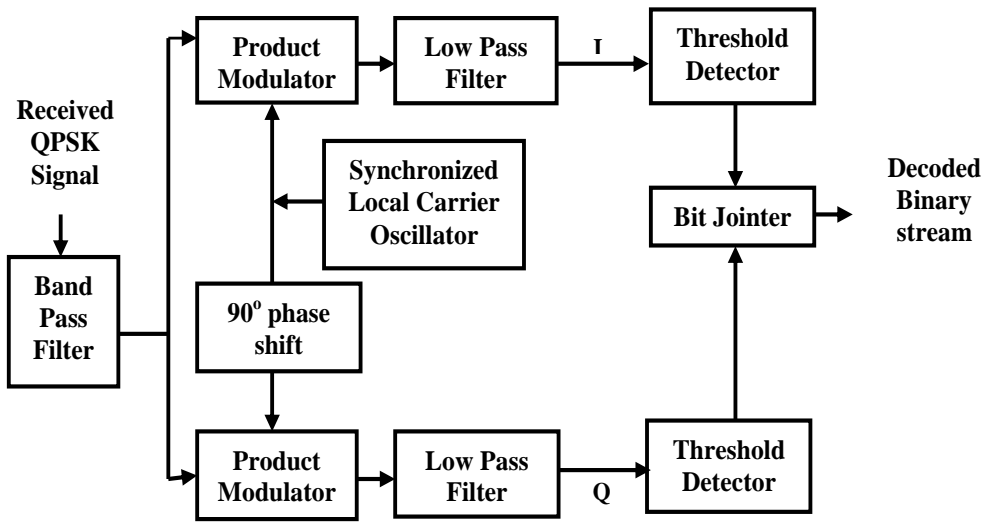

Figure 4 Quadrature Phase shift Keying Demodulator.

## 2.5. Quadrature Amplitude Modulation

Quadrature amplitude modulator comprises of mixer, oscillator and adder as shown in Figure 5 QAM doubles the effective bandwidth by combining two amplitude modulated signals in to one channel. In this scheme, the two high-frequency signals with a phase of 90º between them are amplitude modulated with the two information data sequence known as in-phase(I) and the quadrature-phase (Q) to generate the required output.

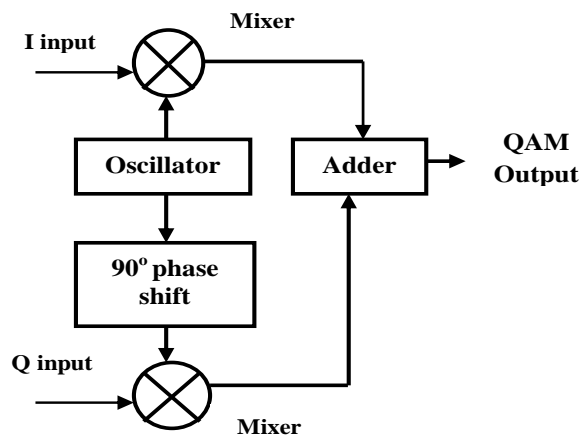

Figure 5. Quadrature Amplitude Modulation.

## 2.6. Quadrature Amplitude Demodulation

The QAM demodulator shown below in Figure 6 utilizes splitter, oscillator and mixer. It does the reverse operation of the modulator. The modulated input signal is fragmented into two separate sequences and each sequence multiplied with a carrier signal generated by the local oscillator using mixer to render in-phase and quadrature components respectively.

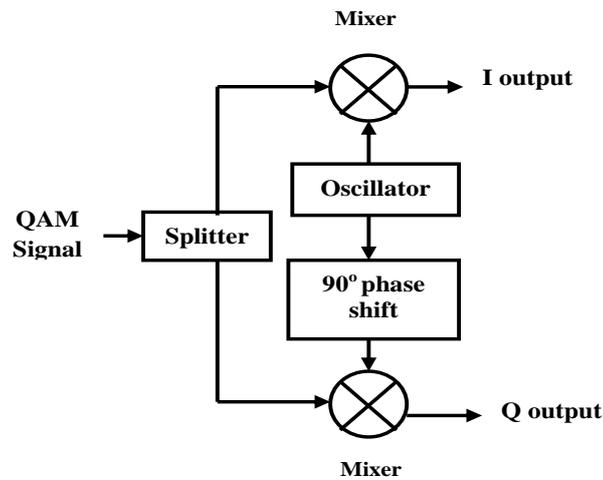

Figure 6. Quadrature Amplitude Demodulation.

## 2.7. Orthogonal Frequency Division Multiplexing (OFDM) System

OFDM is a technique of encoding digital information on various carrier frequencies and used to minimize the spectral width by enhancing the spectral efficiency. In OFDM different carriers are orthogonal to each other and imbrications in the frequency domain to step-up the transmission rate.  The OFDM system consists of serial to parallel converter, parallel to serial converter, symbol mapper, modulator/demodulator and FFT/ IFFT as shown in Figure 7 below. The serial to parallel converter multiplexes high data rate stream into several low bit streams, these streams are converted to BPSK/QPSK/QAM symbols through the symbol mapper. The modulation process of all subcarriers is executed by the IFFT block. The time data from the IFFT block is authorized to the cyclic prefix block for the extension of time samples. The time samples are converted to a serial stream and subsequently converted to the analog domain through digital to analog converter.

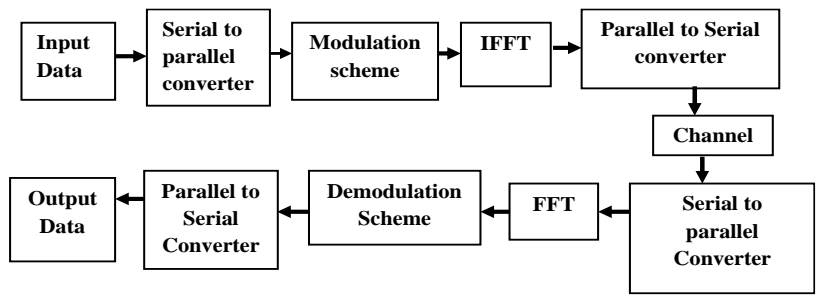

Figure 7. Orthogonal Frequency Division Multiplexing System

FFT is employed to renovate the signals from time domain to frequency domain and inverse Fourier transform is applied to translate the signal back from the frequency domain to the time domain. The cyclic prefix used in frequency division multiplexing schemes acts as a guard band between successive symbols to overcome inter symbol interference (ISI) [23]. The accession of he cyclic prefix contributes robustness to the OFDM signal and it enables the performance to be retained even under higher levels of reflections and multipath propagation.

## 3. RELATED WORK

The quality of the transmitted signal depends on the channel noise and pixel values of the data to be sent [5-8]. The noise increases the bit error rate; in turn, it degrades the signal quality. The signal quality also depends on the processing time [11]. The existing channel estimation algorithm is modified to improve the bit error performance. The bit error analysis is carried out with different modulation techniques with AWGN, Rayleigh and Rician fading channel and concluded that 64QAM increases the design complexity but it is more robust to fading channel [13-14]. Analyzed the performance of different types of equalizers for MQAM and MPSK modulation techniques over a multipath rayleigh fading channel in an outdoor environment to increase the value of the SNR [20-22]. Decision Feedback Equalizer (DFE) shows better performance than linear equalizer. The bit error rate analysis is carried out with various modulation techniques and indicated that rayleigh fading channel is more suitable for multipath propagation of the signal [23-25]. Adaptive modulation is incorporated to improve the signal quality and performance of the wireless system [15-18].

## 4. IMPLEMENTATION

The methodology protruded with setting up the following parameters for execution as shown in Figure 8. The input parameters chosen are

| Input parameters | Numbers |
|---|---|
| FFT Size, No. Of Subcarriers | 64 |
| Cyclic prefix bit | 16 bits |
| Signal to Noise ratio | 0.5 dB, 10 dB, 15dB and 20dB |
| Image pixel size | 1024 X 768 |
| Modulation schemes | BPSK, QPSK, 16QAM and 64QAM |

1. For Image Transmission: Read the image of size 1024*768. Convert the pixel value to a binary stream.

2. For Data Transmission: Data is generated through a random integer generator block.

3. The modulation schemes employed are BPSK, QPSK, 16QAM, and 64QAM. The binary streams are converted to symbols through modulation.

4. Frequency to time-domain conversion is carried out through IFFT. The important block of an OFDM system is the inverse FFT at the transmitting section and FFT at the receiving section. These operations performing linear mappings between N complex data symbols and N complex OFDM symbols result in robustness against fading multipath channel. The data set to be transmitted is X(1), X(2), ..., X(N), where N is the total number of sub-carriers. The discrete-time

$$x(n) = \frac{1}{\sqrt{N}} \sum_{k=0}^{N-1} X(k).e^{j2\pi k \frac{n}{N}}, \qquad n = 0..N-1$$

............... (1)

At the receiver side, the data is recovered by performing FFT on the received signal,

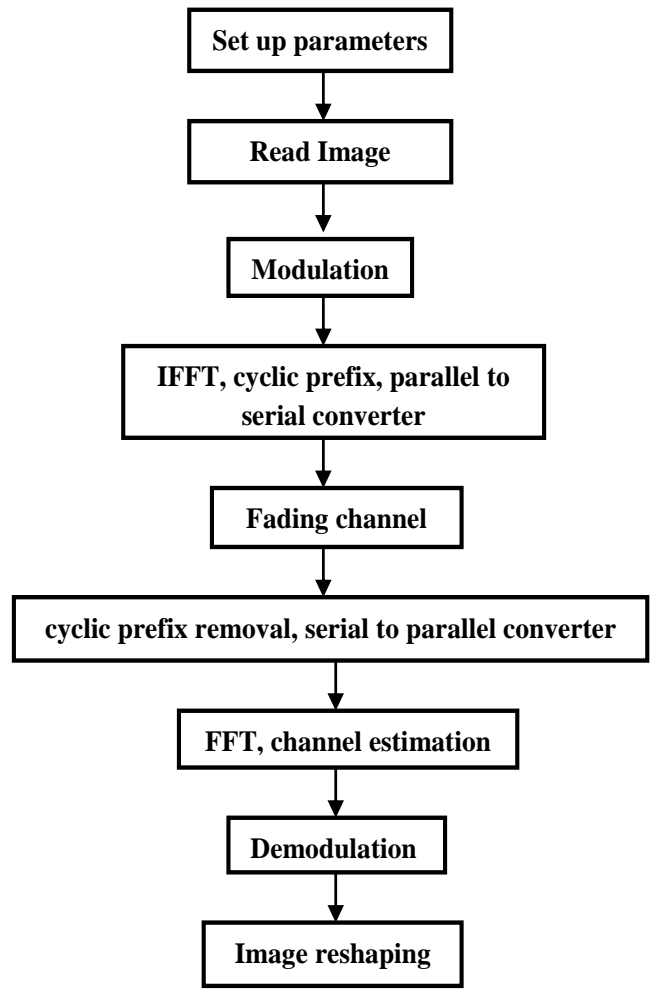

Figure 8. Process flow incorporated for image transmission using OFDM.

$$Y(k) = \frac{1}{\sqrt{N}} \sum_{n=0}^{N-1} x(n).e^{-j2\pi k \frac{n}{N}}, \qquad k = 0..N-1$$

…(2)

5. The time converted signal is cyclic prefixed. Cyclic prefix protects the OFDM signal from inter-symbol interference. The subcarriers generated are represented as shown in the below equation. Figure 9 shows the addition and removal of cyclic prefix in OFDM path.

$$s(k) = \begin{cases} x(k+N) & -M \le k < 0 \\ x(n) = \frac{1}{\sqrt{N}} \sum_{k=0}^{N-1} X(k).e^{j2\pi k \frac{n}{N}} & 0 \le k < N-1 \end{cases}$$

………(3)

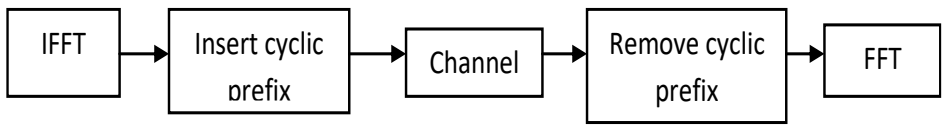

Figure 9. Insertion and removal of the cyclic prefix.

6. A single frequency signal, the pilot signal is transmitted over a communication system for synchronization purpose

7. The parallel data from the output of the IFFT is converted to serial bitstreams using parallel to serial converter. The multiplexer converts the parallel stream of data to a serial data sequences.

8. The output of the parallel to serial converter is fed to the channel input. The serial data streams are passed through the fading channel. The signal is imparted through a fading channel by applying the effect of multipath fading.

9. After passing the signal through multipath fading, the cyclic prefix is removed and the signal is again converted from the time domain to frequency domain by FFT process.

10. After removing the cyclic prefix, the data bits are applied to FFT block, the FFT converts time domain signal back to frequency domain signal. The serial data stream is converted into parallel bits using serial to parallel conversion. The serial data stream is de-multiplexed to parallel data lines. The least-square channel estimation technique removes the effect of multipath fading.

11. The channel estimation is performed using the least square method to eliminate the effect of multipath fading. The frequency response of the channel has to be estimated to invert the effect of non-selective fading on each subcarrier. The transmitter and receiver determine the channel gain at each tone of OFDM symbols. Since the channel transfer function is not changing very rapidly, block type channel estimation is developed as shown in figure 10, in which the pilot tones are inserted into all of the subcarriers of OFDM symbols.

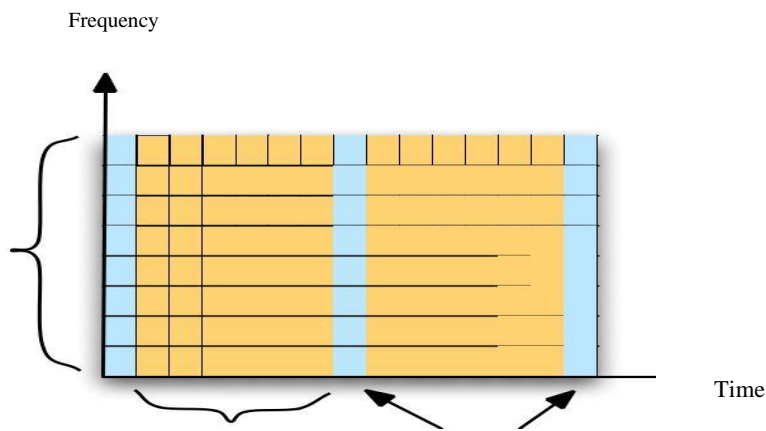

Figure 10: Block type of channel estimation

If the channel is constant during the block, there will be little channel estimation error, since the pilots are sent at all carriers. The pilots are inserted to all subcarriers with a specific period and extracted after DFT block. The estimated channel H(n) is then obtained in a channel estimation block. The received signal can be presented as

$$Y(n) = X(n)H(n) + E(n) \qquad n=1,2,...,N \qquad (4)$$

where E(n) is zero mean noise independent both in time and in frequency due to the linear properties of FFT. The method used to implement the channel estimation is based on the forgetting factor technique:

$$\hat{H}(n) = \alpha(L,n)/\beta(L,n) \qquad n=1.........127 \qquad ............................... \qquad (5)$$

Where

$$\alpha(n,L) = \lambda Xp(n,L)Yp(n,L)+ (1-\lambda)\alpha(n, L-1) \quad ..................................................... \qquad (6)$$

$$\beta(n,L)= \lambda Xp(n,L)Xp(n,L)+ (1-\lambda)\beta(n, L-1) \quad ..................................................... \qquad (7)$$

Where Xp(n, l) is the transmitted pilot, Yp(n, l) is the received pilot, $\lambda$ is the forgetting factor, L is the index of the current pilot frame and n is the sub-channel index. The estimated transmitted signal X(n).

$$\hat{X}(n) = Y(n)/\hat{H}(n) \quad ........................................................................................... \qquad (8)$$

The demodulation performs the reverse operation of the transmitter; the symbols are converted back to binary stream through demodulation. The binary streams are converted back to the original pixel values. The results are analyzed based on BER and SNR.

## 5. RESULTS AND DISCUSSION

The results are discussed in this section. A transmission link is established using BPSK, DBPSK, QPSK, and QAM with OFDM as depicted in Figure 11, 12, 13, 14, 15 and 16 correspondingly.

### 5.1. Transmission link framed with BPSK, DBPSK, QPSK, and QAM

A transmission link utilizing BPSK with OFDM is as shown in Figure 11. The random integer generator block generates random data. The BPSK modulator modulates the input using BPSK. IFFT performs frequency to time conversion and the channel is chosen as AWGN. FFT performs time to frequency conversion. IFFT, AWGN, and FFT represent a basic OFDM system. The BPSK demodulator demodulates the data. The error rate calculator calculates BER and displays it in the display block. The bit error rate is analyzed for 5dB, 10dB and 15dB of Eb/No. The error rate at 5dB, 10dB and 15dB of Eb/No without channel estimation is 0.4948, 0.4964 and 0.4938 respectively. The error rate is 0.099, 0.033 and 0.0072 with channel estimation at Eb/No of 5dB, 10dB, and 15dB respectively. The error rate is lessened with channel estimation along with BPSK and OFDM. The bit error rate is a fundamental parameter for quantifying the performance of a wireless or wired data channel. The bit error rate is the percentage of bits that have errors relative to the total number of bits transmitted [24]. As shown in Figure 12, as the signal to noise ratio increases, the bit error rate decreases.

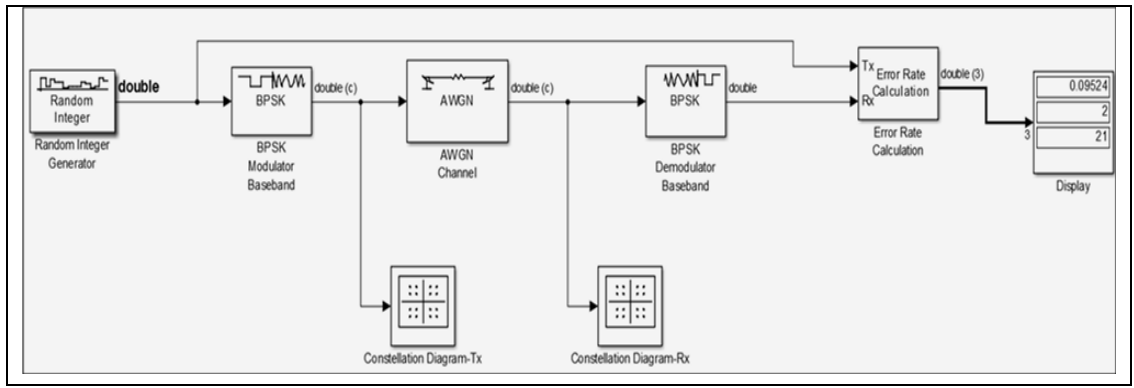

Figure 11. Transmission link framed with BPSK and OFDM.

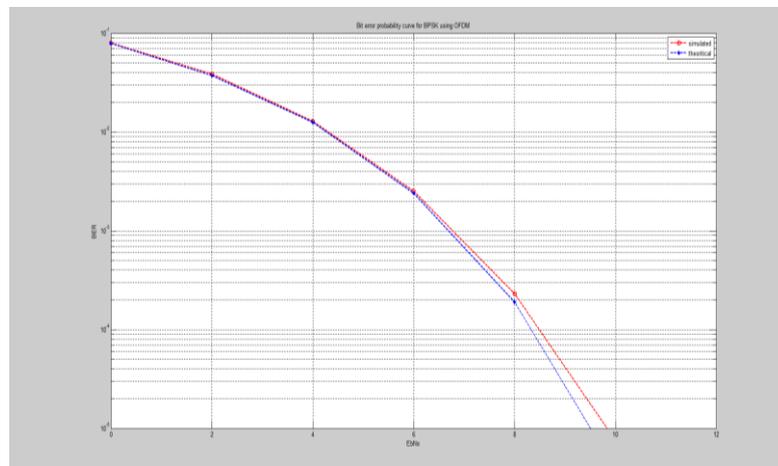

Figure 12. BER vs. SNR of BPSK with OFDM in AWGN

As shown in Figure 13, error analysis with rayleigh fading channel, a higher bit error rate is noticed in comparison with the AWGN channel.

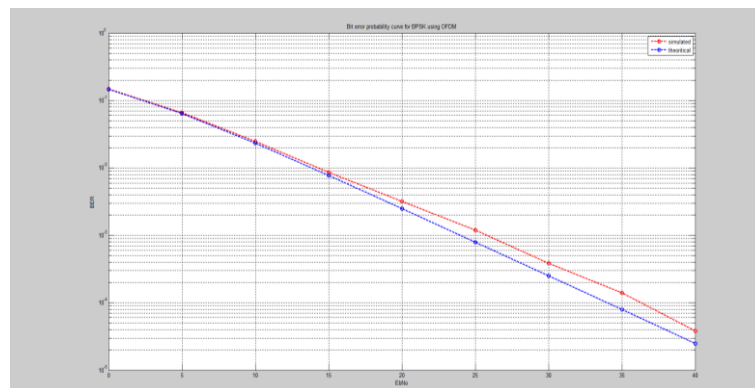

Figure 13. BER vs. SNR of BPSK with OFDM in Rayleigh fading channel

Comparing figures 12 and 13 both show that as SNR increases BER decreases. The above two figures show that the fading channel has more effect on BER than the normal AWGN channel. The BER of BPSK modulation in the AWGN channel with OFDM is low compared to that of BPSK modulation in the rayleigh fading channel, for SNR=0 dB, BER of BPSK in AWGN channel with OFDM=0.078, BER of BPSK in rayleigh fading channel with OFDM=0.1464. Data errors can be minimized by increasing the strength of the signal and by choosing proper error correction techniques. A transmission link using DBPSK with OFDM is as shown in Figure. 14. The bit error rate is analyzed for 5dB, 10dB and 15dB of Eb/No with and without channel estimation. The error rate is compared and observed that the bit error rate is minimized by applying channel estimation.

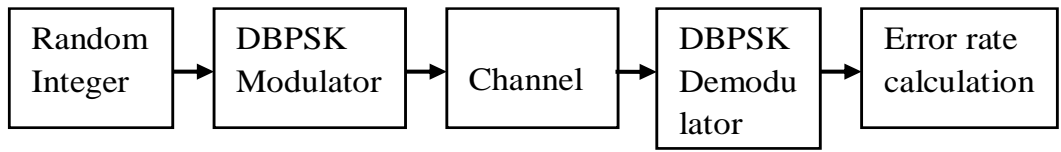

Figure 14. Transmission link framed with DBPSK and OFDM.

A transmission link using QPSK with OFDM is as shown in Figure.15. The bit error rate is examined for 5dB, 10dB and 15dB of Eb/No with and without channel estimation. Without channel estimation the error rate at 5dB, 10dB and 15dB is 0.4987, 0.4985 and 0.4987 and with channel estimation is 0.26, 0.19 and 0.1435 correspondingly. The error rate is compared and observed that the bit error rate is minimized by employing channel estimation. The random integer generator block generates random data. The QPSK modulator modulates the input using QPSK. IFFT performs frequency to time conversion and the channel is chosen is AWGN. FFT performs time to frequency conversion. IFFT, AWGN, and FFT represent a basic OFDM system. The QPSK demodulator demodulates the data. The error rate calculator calculates BER and displays it in the display block.

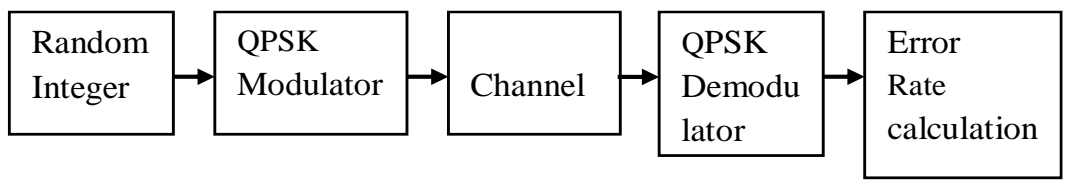

Figure 15. Transmission link framed with QPSK and OFDM.

A transmission link expending QAM with OFDM is as shown in Figure 16. The bit error rate is analyzed for 16QAM at 5dB, 10dB and 15dB of Eb/No with and without channel estimation. The error rate without channel estimation at 5dB, 10dB and 15dB is 0.4965, 0.4957 and 0.4953, and with channel estimation is 0.39, 0.34 and 0.3241 respectively. The bit error rate is analyzed without channel estimation for 64QAM at 5db, 10dB and 15dB correspondingly is 0.4983, 0.4982 and 0.4979 and with channel estimation is 0.43, 0.38 and 0.3762. It is noticed that the error rate derogated with channel estimation.

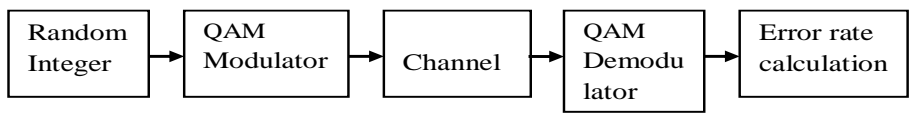

Figure 16. Transmission link framed with QAM and OFDM.

## 5.2. Image outputs

The outcomes obtained for image transmission using BPSK, QPSK, 16QAM, and 64QAM over the AWGN channel with OFDM are represented in Fig. 17, 18, 19 and 20 below.  As shown in Figure 17, an image in the left hand side constitutes the constellation diagram of BPSK and transmitted image of pixel size 1024 X 768.  The right side of the Fig.17 depicts the received signal and its constellation. It has seen that the many constellation points are detected at the receiver, where as the transmitted constellation has only two points. The received signal is noisier compared to the transmitted signal.

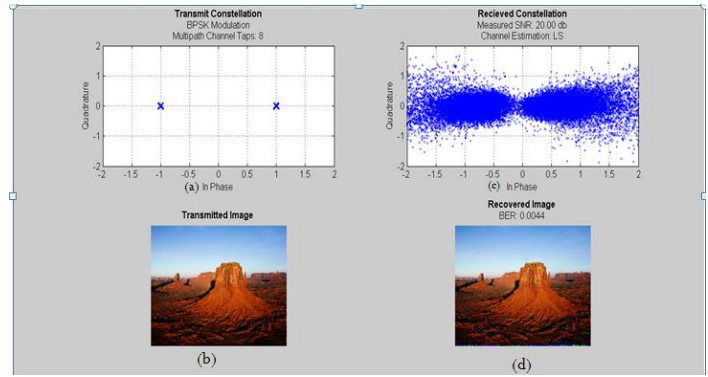

Figure 17. Image Transmission using BPSK with OFDM.

Figure 18 designates the transmitted and received image using QPSK with OFDM in AWGN in SIMULINK.   The transmitted image of size 1024 X 768 and its constellation has four points are exhibited on one side of the image. The received constellation and received image is exhibited on the other side.  The received constellation consists of numerous points as though the transmitted constellation of BPSK consists of only four points and it is scattered compared to transmitted BPSK constellation. The received image is more corrupted than BPSK received image as the number of bits transmitted is more per symbol in QPSK as compared to BPSK.

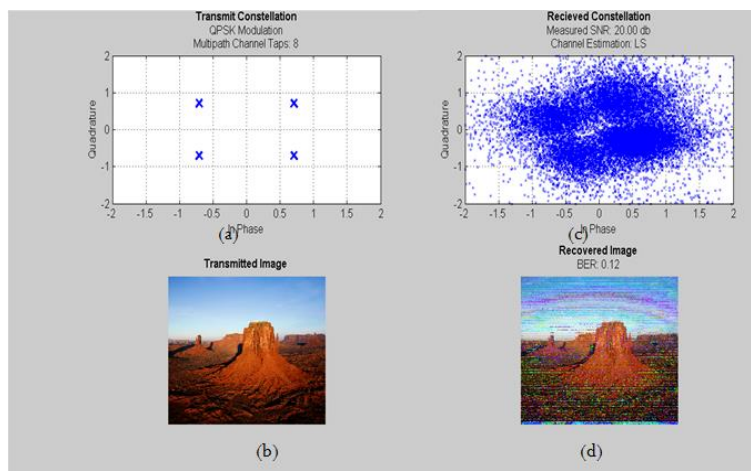

Figure 18. Image Transmission using QPSK with OFDM.

Figure 19 demonstrates the transmitted and received image using 16QAM with OFDM over AWGN channel in SIMULINK.

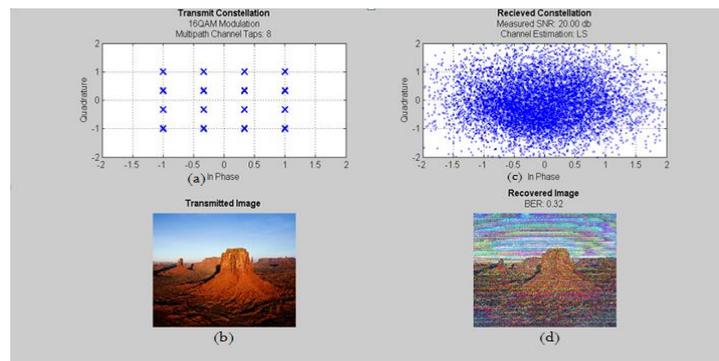

Figure 19. Image Transmission using 16QAM with OFDM.

The constellation diagram and transmitted image of pixel size 1024 X 768 is pictured on one side. The received constellation and received image on the other side. As seen from the received constellation diagram, the concentration is maximum around the region of the transmitted constellation and it is scattered compared to BPSK and QPSK. The received image is extensively corrupted than BPSK and QPSK received images as additional bits per symbol are transmitted. Figure 20, demonstrates the transmitted and received image employing 64QAM with OFDM. The constellation diagram of 64QAM and transmitted image of pixel size 1024 X 768 is revealed on the left side and the received constellation and received image is corresponded on the right side of Fig. 20. As ascertained that several constellation points scattered around the center are observed in 64QAM received constellation than BPSK, QPSK, and 16QAM. The received image of 64QAM is highly corrupted matched to BPSK, QPSK and 16QAM received images as the more number of bits transmitted per symbol.

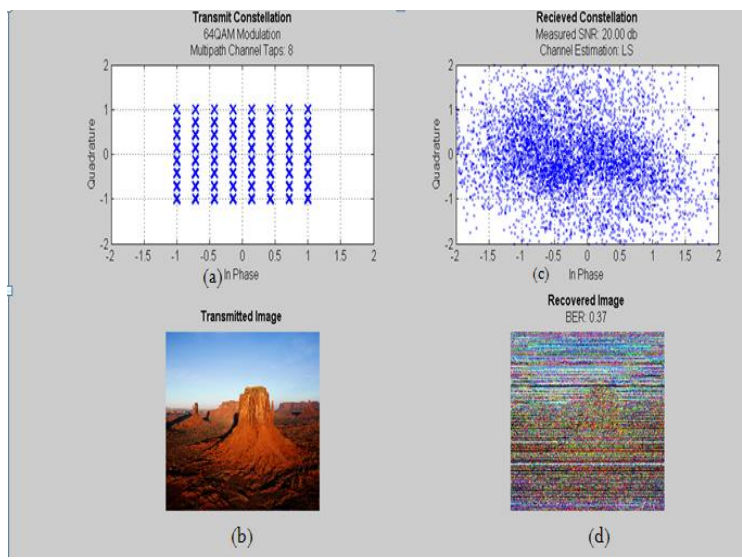

Figure 20. Image Transmission using 64QAM with OFDM.

## 6. COMPARISONS

The results prevailed for transmission link with digital modulation techniques BPSK, QPSK and QAM with and without channel estimation are discussed in this segment.

### 6.1. BER vs. SNR Comparison of BPSK with and without Channel Estimation

Table 1 represents the comparability of bit error rate vs. signal to noise ratio of BPSK modulation with and without channel estimation. The significant part of the mobile wireless channel is the channel estimation. The proper channel estimation technique improves the performance of the system. The channel state information informs about the signal propagation from the transmitter to the receiver and intimates the influence of the fading, scattering and power decay on the signal. The channel estimation method employed here is the least square method to minimize the error rate that occurs due to the effect of multipath fading. The error rate with and without channel estimation is analyzed, compared and commented that the bit errors can be decreased by applying channel estimation technique.

Table 1. BER vs. SNR comparison of Image Transmission using BPSK with and without channel estimation.

| SNR (dB) | Without Channel Estimation | With Channel Estimation |
|----------|---------------------------|-------------------------|
| 5 | 0.4948 | 0.099 |
| 10 | 0.4964 | 0.033 |
| 15 | 0.4938 | 0.0072 |

### 6.2. Comparison of Bit Error Rate vs. Signal to Noise Ratio using QPSK with and without Channel Estimation

Table 2 represents the comparability of BER values of QPSK modulation with and without channel estimation. The channel estimation method incorporated here is the least square method. When no channel estimation is performed, the BER value for the corresponding SNR value is higher compared to that with channel estimation.

Table 2. BER vs. SNR comparison of Image Transmission using QPSK with and without channel estimation.

| SNR (dB) | Without Channel Estimation | With Channel Estimation |
|----------|---------------------------|-------------------------|
| 5 | 0.4987 | 0.2600 |
| 10 | 0.4985 | 0.1900 |
| 15 | 0.4983 | 0.1435 |

### 6.3. Comparison of Bit Error Rate vs. Signal to Noise Ratio using 16QAM with and without Channel Estimation

Table 3 represents the comparison of BER values of 16QAM modulation with and without channel estimation. The BER value for the corresponding SNR value is tabulated and compared. It is detected that bit errors can be minimized by employing channel estimation technique.

Table 3. BER vs. SNR comparison of Image Transmission using 16QAM with and without channel estimation.

| SNR (dB) | Without Channel Estimation | With Channel Estimation |
|---|---|---|
| 5 | 0.4965 | 0.3900 |
| 10 | 0.4957 | 0.3400 |
| 15 | 0.4953 | 0.3241 |

## 6.4. Comparison of Bit Error Rate vs. Signal to Noise Ratio using 64QAM with and without Channel Estimation

Table 4 represents the comparability of BER values of 64QAM with and without channel estimation. The channel estimation method applied is least square method. The error rate is comparatively less with channel estimation method. As seen from Tables 1, 2, 3 and 4 comparing the BER values of the digital modulation techniques such as BPSK, QPSK, 16-QAM, and 64-QAM, the BPSK modulation has the minimum BER value compared to all the other techniques. In BPSK modulation the number of bits transmitted per symbol is two, which is least compared to other modulation techniques. It is concluded from the outcomes that the bit error increases with an increase in the number of bits transmitted per symbol as indicated in the bar graph below in Fig.17.

Table 4. BER vs. SNR comparison of Image Transmission using 64QAM with and without channel estimation.

| SNR (dB) | Without Channel Estimation | With Channel Estimation |
|---|---|---|
| 5 | 0.4983 | 0.4300 |
| 10 | 0.4982 | 0.3800 |
| 15 | 0.4979 | 0.3762 |

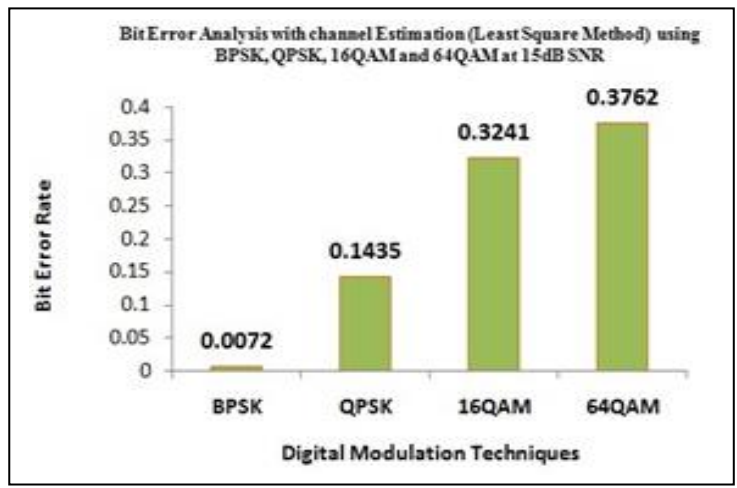

Figure 21. Comparison of BER Analysis with channel estimation using BPSK, QPSK, 16QAM, and 64QAM at 15dB

Table5. BER vs. SNR comparison of BPSK in AWGN and Rayleigh channel

| SNR (dB) | BPSK (AWGN) | BPSK (Rayleigh) |
|----------|-------------|-----------------|
| 0 | 0.07865 | 0.1464 |
| 2 | 0.03751 | 0.1085 |
| 4 | 0.0125 | 0.07714 |

Table 5 represents the comparison of BER values of the digital modulation techniques such as BPSK in AWGN with OFDM, BPSK in Rayleigh Fading channel with OFDM. The results show that the BER values of BPSK in AWGN is less than that of the rayleigh fading channel for the corresponding SNR.

# 7. CONCLUSION

This work main concentrate on the transmission of an image signal over a noisy channel and calculation of bit error rate Vs. signal to noise ratio. The data and image signal are modulated with BPSK, QPSK, 16QAM and 64QAM schemes and transmitted over a noisy channel. The results obtained are compared with and without channel estimation and remarked that the bit error rate is minimum with OFDM and channel estimation comparatively without channel estimation. The received image of the 64QAM is more corrupted than in comparison with the received image of the BPSK, QPSK, and 16QAM. The bit error rate increases with the order of modulation. The same set up can be enhanced for 128QAM, 256 QAM, 512 QAM and 1024 QAM with Multiple Input and Multiple Output (MIMO) with adaptive modulation techniques.

## ACKNOWLEDGMENTS

I would like to thank JSS academy of Technical education for lab support to execute this work. I would like to thank my family members for bearing me while doing this work.

**AUTHORS**

**Dr. Usha S.M.** from Bengaluru- Karnataka, India obtained B.E (Electronics & Communication Engineering) degree from Mysore University in the year 2000. M.Tech. in VLSI Design and Embedded Systems from VTU Belgaum in 2011 and awarded Ph.D.  in Optimization and Performanc e Analysis of  Digital Modulators form VTU Belgaum in the year 2017.  She is currently working as Associate Professor at JSS Academy of Technical Education, Bengaluru. Karnataka, India. She is a member of Professional bodies such as IEEE, ISTE, and MIE.

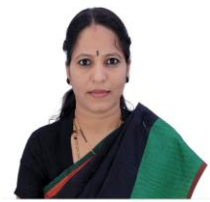

**Mr. Mahesh H.B**. from Bengaluru-Karnataka, India obtained Computer Science B.E, a degree from Mysore University in the year 1996, M.Tech in Networking & Internet Engineering from VTU in the year 2004. Currently, working as an Assistant Professor at PES University, Bengaluru, and Karnataka, India. He is a member of professional bodies such as IEEE, CSE.

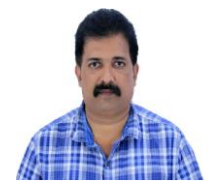